\documentclass[aps,prd,twocolumn,groupedaddress,showpacs]{revtex4}
\RequirePackage{amsmath2000}

\newcommand{\Deqn}[1]{{Eq.~(\ref{#1})}}
\newcommand{\x}{x}
\newcommand{\z}{z}
\newcommand{\tail}{{\text{tail}}}
\newcommand{\self}{{\text{self}}}
\newcommand{\dir}{{\text{dir}}}
\newcommand{\ret}{{\text{ret}}}
\newcommand{\adv}{{\text{adv}}}
\newcommand{\rad}{{\text{rad}}}
\newcommand{\sym}{{\text{sym}}}
\newcommand{\R}{{\text{R}}}
\newcommand{\SSS}{{\text{S}}}
\newcommand{\gB}{\bar{g}}
\newcommand{\hB}{\bar{h}}
\newcommand{\p}{\prime}
\newcommand{\q}{{q}}
\newcommand{\ddd}{\mbox{d}}

\begin{document}

\title{Self force via a Green's function decomposition}
\author{Steven Detweiler}
\author{Bernard F. Whiting}
\affiliation{Department of Physics, PO Box 118440, University of Florida,
          Gainesville, FL 32611-8440}
\date{November 12, 2002}

\begin{abstract}
The gravitational field in a neighborhood of a particle of small mass
$\mu$ moving through curved spacetime is naturally decomposed into
two parts each of which satisfies the perturbed Einstein equations
through $O(\mu)$. One part is an inhomogeneous field which looks like
the $\mu/r$ field tidally distorted by the local Riemann tensor. The
other part is a homogeneous field that completely determines the self
force of the particle interacting with its own gravitational field,
which changes the worldline at $O(\mu)$ and includes the effects of
radiation reaction. Surprisingly, a local observer measuring the
gravitational field in a neighborhood of a freely moving particle
sees geodesic motion of the particle in a perturbed vacuum geometry
and would be unaware of the existence of radiation at $O(\mu)$.
 In the light of all previous work this is quite an
unexpected result.

\end{abstract}

\pacs{ 04.25.-g, 04.20.-q, 04.25.Nx, 04.30.Db}

\maketitle

\section{Introduction}
The principle of equivalence in General Relativity implies that a freely
moving particle, of infinitesimal mass and size, follows a geodesic of
spacetime\cite{MTW}.
 For a small but finite mass $\mu$, the particle perturbs the spacetime
geometry at $O(\mu)$.  The interaction of the particle with this
perturbation of the metric is called the ``self force'' which changes the
worldline at $O(\mu)$ and includes the effects of radiation reaction.

A local observer measuring the gravitational field in a neighborhood of the
particle, with no a priori knowledge of the background geometry, sees a
combination of the background metric plus its $O(\mu)$ perturbation caused
by the particle. Local measurements, in a neighborhood of the particle,
cannot distinguish these two specific contributions. Nevertheless, we show
below that the combined metric in the neighborhood of the particle can be
uniquely decomposed into two distinct parts: $(i)$ the tidally distorted
$\mu/r$ piece of the perturbation of the metric, which is an inhomogeneous
solution of the perturbed Einstein equations at $O(\mu)$, and $(ii)$ the sum
of the background metric and the remainder of the metric perturbation, which
together are a homogeneous solution of the Einstein equations through
$O(\mu)$.

The self force is shown to result in geodesic motion of the particle,
through $O(\mu)$, in the homogeneous perturbed metric, $(ii)$ above. With
only local measurements, the observer has no means of distinguishing the
homogeneous perturbation from the background metric through $O(\mu)$. As the
particle moves along a geodesic of this perturbed vacuum geometry, the
observer sees no effect which one would be compelled to interpret as
radiation reaction at $O(\mu)$. In this precisely defined fashion, we extend
the understanding of the principle of equivalence in General Relativity: The
orbit of the particle is a geodesic through $O(\mu)$, in a locally
measurable, vacuum gravitational field.

\section{Historical background}
In Dirac's \cite{Dirac38} analysis of the self force for the
electromagnetic field of a particle in flat spacetime, he used the
conservation of the stress-energy tensor inside a narrow world tube
surrounding the particle's worldline to derive equations of motion with
radiation reaction effects included. DeWitt and Brehme
\cite{DeWittBrehme60} extended this approach to allow for curvature in
spacetime. Mino, Sasaki and Tanaka \cite{Mino97} generalized it to include
the gravitational self force as well.
 Quinn and Wald\cite{QuinnWald97} and Quinn\cite{Quinn00} have obtained
similar results by pursuing an axiomatic approach for the gravitational,
electromagnetic and scalar field self forces.

On a formal level, these analyses provide a coherent and consistent
treatment of the self force.
 Analysis begins with a given worldline $\Gamma$, described by
$z^a(\tau)$, which obeys the Lorentz force law through the fixed
background gravitational and electromagnetic fields.  The self force gives
the actual worldline an acceleration away from $\Gamma$, and it is the
mass times this acceleration which it is wished to determine.

Towards this end, first the retarded field is obtained in terms of the
corresponding Green's function. Traditionally the Green's function
 has been decomposed into ``direct'' and ``tail'' parts.
The resulting ``direct'' part of the field at a point $\x$ is determined
completely by sources on the past null cone of $\x$; in flat spacetime
this is the Li\'enard-Wiechert potential for electromagnetism. The
curvature of spacetime allows for an additional contribution from sources
within the past null cone, and this is the ``tail'' part. The self force
on a particle is then composed of two pieces: The first piece comes from
the direct part of the field and the acceleration of $\Gamma$ in the
background geometry; in flat spacetime this is the Abraham-Lorentz-Dirac
(ALD) force. The second piece comes from the tail part of the field and is
present in curved space even if $\Gamma$ is a geodesic.  While the
decomposition of the field into the direct and tail parts has been useful
for describing the self force, neither of these parts individually is a
solution of the field equation.

In flat spacetime Dirac \cite{Dirac38} decomposed the retarded
electromagnetic field into two parts: $(i)$ the ``mean of the advanced and
retarded fields'' which is an inhomogeneous field resembling the Coulomb
$q/r$ piece of the scalar potential near the particle, and $(ii)$ a
``radiation'' field, his Eqs.~(11) and (13), which is a homogeneous solution
of Maxwell's equations.  He described the self force as the interaction of
the particle with the radiation field, a well-defined vacuum Maxwell field.

Previous descriptions of the self force in curved spacetime
\cite{DeWittBrehme60, Mino97, QuinnWald97,Quinn00} reduce to Dirac's result
in the flat spacetime limit. And they provide clear, adequate expressions
for the self force.  However, they do not share the physical simplicity of
Dirac's analysis where the force is described entirely in terms of an
identifiable, vacuum solution of the field equations. Indeed, for an
electromagnetic field, the vector potential $A_a^{\text{self}}$
[\textit{cf.} \Deqn{psiself} for the corresponding scalar field expression]
might be said to be responsible for the self force. But generally
$A_a^{\text{self}}$ does not satisfy \Deqn{MaxEqn} below, and the current
density $J^a$, which would result from the application of the operator on
the left hand side of \Deqn{MaxEqn} to $A_a^{\text{self}}$, would have a
non-vanishing charge distribution in the vicinity of the particle.
Furthermore, $A_a^{\text{self}}$ is non-differentiable at the location of
the particle with four-velocity $u^a$ if $(R_{ab}-\frac{1}{6}g_{ab}R) u^b
\ne 0$ there.
 Similar statements hold for scalar and gravitational fields.

In this letter we present a curved space generalization of Dirac's scheme.
We find that, near the worldline $\Gamma$, even in a curved geometry, the
retarded field can be decomposed into two parts such that the first,
$A_a^\SSS$, is an inhomogeneous solution to the field equation with a point
source, similar to the ``mean of the advanced and retarded fields,'' while
the second, $A_a^\R$, is a homogeneous solution which yields a complete
description of the self force. These parts are related to but distinct from
the usual direct and tail parts. Scalar and gravitational fields are
analyzed similarly.

Below, we first focus on the scalar field case, and then briefly describe
our results for the electromagnetic and gravitational fields.

\section{Scalar field}
The force on a point particle with charge $q$ moving in a scalar field may
be deduced from (see discussion in \cite{Quinn00})
\begin{equation}
   F^a = q\nabla^a \psi ,
\label{scalarforce}
\end{equation}
where the derivative of the field is to be evaluated at the location of
the particle.  Usually it is implicit that $\psi$ does not include the
field from the particle itself but is composed only of the ``external''
field.  The description of the self force is not so straightforward.
However, one can introduce a quantity $\psi^\self$, as in our
\Deqn{psiself}, which when substituted into the right hand side of
\Deqn{scalarforce} formally yields the scalar-field self force as
described by Quinn \cite{Quinn00}. Similar expressions for the force in
terms of the derivatives of the field are given in \Deqn{emforce} and
\Deqn{gravforce} for electromagnetic and gravitational fields.

The scalar field equation
\begin{equation}
   \nabla^2 \psi \equiv \nabla^a \nabla_a \psi = -4\pi\rho
\label{DelPsi}
\end{equation}
is formally solved in terms of a Green's function,
\begin{equation}
  \nabla^2 G(x,z) = - (-g)^{-1/2} \delta^4(x-z) .
\label{Greenseqn}
\end{equation}
The source function for a point charge moving along a worldline
$\Gamma$, described by coordinates $z^a(\tau)$, is
\begin{eqnarray}
   \rho(x) &=& \q \int (-g)^{-1/2} \delta^4(x-z(\tau)) \,\ddd\tau ,
\label{rhox}
\end{eqnarray}
where $\tau$ is the proper time along the worldline of the particle with
scalar charge $\q$. The scalar field of this particle is
\begin{equation}
  \psi(\x) = 4\pi \q \int G[\x,z(\tau)] \,\ddd\tau .
\end{equation}

A symmetric scalar field Green's function is derived from the Hadamard
form to be
\begin{equation}
  G^\sym(\x,\z) = \frac{1}{8\pi} \left[ u(\x,\z) \delta(\sigma)
         - v(\x,\z) \Theta(-\sigma) \right] ,
\label{greenSSym}
\end{equation}
where $u(\x,\z)$ and $v(\x,\z)$ are bi-scalars, the properties of which
are described by DeWitt and Brehme \cite{DeWittBrehme60}. They are
determined by a local expansion in the vicinity of $\Gamma$, and are
symmetric under interchange of $\x$ and $\z$. $\sigma$ is half of the
square of the distance measured along the geodesic from $\x$ to $\z$ with
$\sigma<0$ for a timelike geodesic, and $\sigma=0$ on the past and future
null cones of $\x$.
 The expansions for the bi-scalars $u(\x,\z)$ and $v(\x,\z)$
are known to be convergent within a finite neighborhood of $\Gamma$ if the
geometry is analytic\cite{Hadamard23}.
 The $\Theta(-\sigma)$ guarantees that only when $\x$ and $\z$ are timelike
related is there a contribution from $v(\x,\z)$.
 The terms in any Green's function containing $u$ and $v$ are frequently
referred to as the ``direct'' and ``tail'' parts, respectively.

The direct part of $G^\sym(\x,\z)$, has support only on the null cone of
$\x$, and the resultant direct part of the field for a particle moving
along $\Gamma$ is
\begin{equation}
  \psi^\sym_\dir(\x) = \left[\frac{\q
  u(\x,\z)}{2\dot{\sigma}}\right]_{\tau_\ret}
                      + \left[\frac{\q
                      u(\x,\z)}{2\dot{\sigma}}\right]_{\tau_\adv} ,
\label{psidir}
\end{equation}
where $\dot{\sigma} = \ddd\sigma (x,z(\tau))/\ddd\tau$, and
${\tau_\ret}$ and ${\tau_\adv}$ refer to the proper time of the
intersection of $\Gamma$ with the past and future null cones of $\x$,
respectively.

The tail part of $G^\sym(\x,\z)$, has support within both the past and
future null cones of $\x$, and the resultant tail part of the field for a
particle moving along $\Gamma$ is
\begin{eqnarray}
  \psi^\sym_\tail(\x) &=&
       - \frac{\q}{2} \left( \int_{-\infty}^{\tau_\ret}
                  +  \int_{\tau_\adv}^{\infty} \right)  v(\x,\z) \,\ddd\tau .
\label{xpsiSother}
\end{eqnarray}

DeWitt and Brehme \cite{DeWittBrehme60} use local expansions in the
vicinity of $\Gamma$ to show that \footnote{Our convention
$2\nabla_{[a}\nabla_{b]} \xi_c = R_{abc}{}^d \xi_d$ agrees with that in
references \cite{QuinnWald97,Quinn00} and is the opposite of that used in
\cite{DeWittBrehme60,Mino97}.}
\begin{equation}
  u(\x,\z) = 1 +\frac{1}{12}R_{ab} \nabla^a\sigma \; \nabla^b\sigma +
   O(r^3), \quad\x\rightarrow\Gamma ;
\label{uexpand}
\end{equation}
$r$ is the proper distance from $\x$ to $\Gamma$ measured along the
spatial geodesic which is orthogonal to $\Gamma$.
 They also show that the symmetric bi-scalar $v(\x,\z)$ is a solution of
the homogeneous wave equation,
\begin{equation}
  \nabla^2 v(\x,\z) = 0,
\label{del2v}
\end{equation}
and that for $\x$ close to $\Gamma$,
\begin{equation}
   v(\x,\z) = - \frac{1}{12} R(\z) + O(r), \quad\x\rightarrow\Gamma .
\label{vexpand}
\end{equation}

The retarded and advanced Green's functions are derived from
$G^\sym(x,z)$,
\begin{eqnarray}
   G^\ret(\x,\z) &=& 2\Theta[\Sigma(\x),\z] G^\sym(\x,\z)
\nonumber\\
   G^\adv(\x,\z) &=& 2\Theta[\z,\Sigma(\x)] G^\sym(\x,\z) ,
\label{greenRetAdv}
\end{eqnarray}
where $\Theta[\Sigma(\x),\z]= 1 - \Theta[\z,\Sigma(\x)]$ equals 1 if $\z$
is in the past of a spacelike hypersurface $\Sigma(\x)$ that intersects
$\x$, and is zero otherwise. The $G^\rad(\x,\z)$ implicitly used by Dirac
in flat spacetime is
\begin{equation}
   G^\rad(\x,\z) = G^\ret(\x,\z) - G^\sym(\x,\z).
\label{radgreen}
\end{equation}
Note that the fields resulting from two different Green's functions that
each obey \Deqn{Greenseqn} necessarily differ by a homogeneous solution of
\Deqn{DelPsi}.

$G^\ret(\x,\z)$ has reasonable causal structure, and we assume for
simplicity that $\psi^\ret$ is, in fact, the actual field resulting from
the source particle.

\section{Self force}
Careful analyses \cite{DeWittBrehme60,Mino97,QuinnWald97,Quinn00} show
that contributions to the self force result from both the source's
acceleration, if $\Gamma$ is not a geodesic, and from the curvature of
spacetime. For these two distinct possibilities, the self force is a
consequence of the particle interacting with either the direct or the tail
part of its field, respectively. By following the detailed derivations in
\cite{DeWittBrehme60,Mino97,QuinnWald97,Quinn00}, we see that the self
force may be considered to result, via \Deqn{scalarforce}, from the
interaction of the particle with the quantity
\begin{equation}
  \psi^\self=
    - \left[\frac{\q u(\x,\z)}{2\dot{\sigma}}\right]_{\tau_\ret}^{\tau_\adv}
             - \q \int_{-\infty}^{\tau_\ret} v[\x,\z(\tau)] \,\ddd\tau
\label{psiself}
\end{equation}
where, unlike Dirac's radiation field, $\psi^\self$ is not a homogeneous
solution of the field equation.

The first expression in \Deqn{psiself} is finite and differentiable in the
coincidence limit, $\x\rightarrow\Gamma$. When substituted into the right
hand side of \Deqn{scalarforce} this expression provides the curved space
generalization of the ALD force, and local expansions of $u(\x,\z)$ and
$\dot\sigma(\x,\z)$ in \cite{DeWittBrehme60,Mino97,QuinnWald97,Quinn00}
 give the resultant force in terms of the acceleration of
$\Gamma$ and components of the Riemann tensor.

The integral in \Deqn{psiself} comes from the tail part of the Green's
function. Its derivative results, in part, from the implicit dependence of
$\tau_\ret$ upon $x$. Quinn \cite{Quinn00} computes this contribution to
the derivative to be
\begin{multline}
  -q v[\x,\z(\tau_\ret)] \nabla_a\tau_\ret
    =  q \left[ v \dot\sigma^{-1} \nabla_a\sigma \right]_{\tau_\ret} \\
    = - \frac{q R(x)}{12r}(x_a - z_a)
     + O(r),
      \quad\x\rightarrow\Gamma ,
\label{trouble}
\end{multline}
the spatial part of which is not defined when $\x$ is on $\Gamma$. In the
usual self force analysis, one first averages $\nabla_a\psi^\self$ over a
small, spatial two-sphere surrounding the particle, thus removing the
spatial part of \Deqn{trouble}. Then one takes the limit of this average
as the radius of the two-sphere goes to zero, thereby obtaining a well
defined contribution to the self force
\cite{DeWittBrehme60,Mino97,QuinnWald97,Quinn00}.

\section{Homogeneous field} We now provide an alternative expression
for the field responsible for the self force.

Given one Green's function which solves \Deqn{Greenseqn}, a second may be
generated by adding to the first any bi-scalar which is a homogeneous
solution of \Deqn{Greenseqn}. $v(\x,\z)$ is just such a bi-scalar and is
also symmetric, $v(\x,\z) = v(\z,\x)$ \cite{DeWittBrehme60}. Thus, a new
symmetric Green's function is
\begin{eqnarray}
  G^\SSS(\x,\z) &\equiv& G^\sym(\x,\z)  + \frac{1}{8\pi} v(\x,\z)
\nonumber\\
     &=&  \frac{1}{8\pi}\left[ u(\x,\z) \delta(\sigma)
         + v(\x,\z) \Theta(\sigma) \right] ,
\label{greenS}
\end{eqnarray}
which has no support within the null cone.
 We use $G^\SSS(\x,\z)$ only in a local neighborhood of the particle,
and do not depend upon any knowledge of its global existence.
 $G^\SSS(\x,\z)$ does have support on the null cone of $\x$, just as
$G^\sym(\x,\z)$ does, and also outside the null cone, at spacelike
separated points.  The use of $G^\SSS(\x,\z)$ is thus not complicated
by the need for knowledge of the entire past history of the source
and is amenable to local analysis. The corresponding field
\begin{eqnarray}
  \psi^\SSS(\x) &=& \left[\frac{\q u(\x,\z)}{2\dot{\sigma}}\right]_{\tau_\ret}
       + \left[\frac{\q u(\x,\z)}{2\dot{\sigma}}\right]_{\tau_\adv}
\nonumber\\ &&
    {} + \frac{\q}{2} \int_{\tau_\ret}^{\tau_\adv} v(\x,\z) \,\ddd\tau
\label{psiSother}
\end{eqnarray}
is an inhomogeneous solution of \Deqn{DelPsi} just as $\psi^\ret$ is. In
the pioneering spirit of Dirac, it is natural to define
\begin{equation}
  G^\R(\x,\z) \equiv G^\ret(\x,\z) - G^\SSS(\x,\z)
\end{equation}
[\textit{cf.} \Deqn{radgreen}].
 Remarkably, like $G^\ret(\x,\z)$, $G^\R(\x,\z)$ has no support inside the
future null cone. Corresponding to $G^\R(\x,\z)$, we construct
\begin{eqnarray}
  \psi^\R &=& \psi^\ret -  \psi^\SSS
\nonumber\\ &=&
    - \left[\frac{\q u(\x,\z)}{2\dot{\sigma}}\right]^{\tau_\adv}_{\tau_\ret}
\nonumber\\ &&
       {} - \q\left( \int_{-\infty}^{\tau_\ret}
       + \frac{1}{2} \int_{\tau_\ret}^{\tau_\adv}\right) v(\x,\z) \,\ddd\tau
\label{psiR}
\end{eqnarray}
analogous to Dirac's radiation field.
 By construction $\psi^\R$ is a homogeneous solution of \Deqn{DelPsi}
and has
the consequent property that it is smooth in the coincidence limit,
$\x\rightarrow\Gamma$. We note its relation to $\psi^\self$
\begin{equation}
  \psi^\R =  \psi^\self
         - \frac12 \q \int_{\tau_\ret}^{\tau_\adv} v(\x,\z) \,\ddd\tau  .
\label{psidiff}
\end{equation}

Our most significant technical result is, in fact, that the self force is
determined by the particle's interaction with $\psi^\R$, since the integral
term in \Deqn{psidiff} gives no contribution to a self force.  For a field
point $\x$ which is near $\Gamma$, the integrand may be expanded using
\Deqn{vexpand}. The dominant part of the integral from $\tau_\ret$ to
$\tau_\adv$ then brings in a factor of $\tau_\adv-\tau_\ret = 2r + O(r^2)$
times $v(x,x)$, and the integral term of \Deqn{psidiff} is
\begin{equation}
  - q r v(x,x)+O(r^2) = \frac{1}{12}\q r R(\x)
           + O(r^2), \quad\x\rightarrow\Gamma .
\end{equation}
The derivative of $\frac{1}{12}\q r R(\x)$ gives an outward pointing,
spatial unit vector near $\Gamma$; this exactly cancels the troublesome
part of $\nabla_a\psi^\self$ in \Deqn{trouble} which is thus absent from
$\nabla_a\psi^\R$. The derivative of the remainder term $O(r^2)$ is zero
in the limit that $x$ approaches $\Gamma$ and gives no contribution to the
self force. Thus, the self force may be seen to be due exclusively to the
interaction of the particle with $\psi^\R$ via \Deqn{scalarforce}. We
regard this approach as preferable, because $\psi^\R$ is differentiable at
the location of the particle, so that averaging is no longer required in
computing the self force. Even more importantly, $\psi^\R$ is a
homogeneous solution of \Deqn{DelPsi}.

\section{Electromagnetic and gravitational fields} The analysis for the
scalar field is easily generalizable to both electromagnetic and
gravitational fields by the addition of extra indices to $u(\x,\z)$ and
$v(\x,\z)$ to create corresponding bi-vectors and bi-tensors, and by the
introduction of $\gB^{ab^\p}(\x,\z)$, which is the bi-vector of parallel
displacement \cite{DeWittBrehme60}. The primed indices below refer to the
source point $\z$, the unprimed indices to the field point $\x$ as above.
The definitions and relationships for the various Green's functions follow
the same pattern as above and are not repeated below.

For the electromagnetic field, the Lorentz gauge requires
\begin{equation}
  \nabla_a A^{a} = 0 ,
\label{harmonicA}
\end{equation}
so that Maxwell's equations become
\begin{equation}
 \nabla^2 A^a
           - {R^a}_b A^b
      = - 4\pi J^a .
\label{MaxEqn}
\end{equation}
DeWitt and Brehme \cite{DeWittBrehme60} show that
\begin{equation}
  u_{a b^\p}(\x,\z)
        =  \gB_{a b^\p}(\x,\z)
           u(\x,\z) ,
\label{uab}
\end{equation}
and that (as $x\rightarrow\Gamma$)
\begin{equation}
   v_{a b^\p}(\x,\z) = \frac12 \gB_a{}^{c^\p} \Bigl( R_{b^\p c^\p}
             - \frac{1}{6} g_{b^\p c^\p}R\Bigr) + O(r) .
 \label{vab}
\end{equation}
If $(R_{ab}-\frac{1}{6}g_{ab}R) u^b \ne 0$ at the particle, then
$A^\text{self}_a$ is non-differentiable there.

Similar to the scalar field, $A^\SSS_{a}$ is an inhomogeneous
solution of \Deqn{MaxEqn}. That it also satisfies the gauge condition
(\ref{harmonicA}) follows from an argument similar to that after Eqn.
(3.37) in \cite{DeWittBrehme60}.  We have
\begin{equation}
  \nabla^a A^\SSS_a = \int \nabla_{a^\p}(G^\SSS J^{a^\p})\sqrt{-g}\,\ddd^4x^\p,
\end{equation}
assuming that $J^a$ is conserved, where $G^\SSS$ is the scalar
Green's function of \Deqn{greenS}.  That $G^\SSS$ has no support
within the past or future null cone implies that the integral is zero
when written as a boundary integral.

With the definition
\begin{equation}
  A^\R_{a} \equiv A^\ret_{a} - A^\SSS_{a},
\end{equation}
$A^\R_{a}$ gives a homogeneous electromagnetic field. The
electromagnetic self force becomes
\begin{equation}
  F^a = q g^{ac} (\nabla_c A^\R_b - \nabla_b A^\R_c ) \dot z^b .
\label{emforce}
\end{equation}
This combines with the Lorentz force law from the background to determine
the actual worldline of the particle.

For the gravitational field, the harmonic gauge requires
\begin{equation}
  \nabla_a \hB^{ab} = 0 ,
\label{deDonder}
\end{equation}
and, with $R_{ab}=0$, the perturbed Einstein equations are
\begin{equation}
 \nabla^2 \hB_{ab}
           + 2{R_a}^c{}_b{}^d \hB_{cd}
      = - 16\pi T_{ab} ,
\label{EEqnHarm}
\end{equation}
where $\hB_{ab}=h_{ab}-\frac12 g_{ab} h^c{}_c$ is the trace reversed
version of the metric perturbation $h_{ab}$. Mino, Sasaki and Tanaka
\cite{Mino97} show that
\begin{equation}
  u_{abc^\p d^\p}(\x,\z)
        = 2 \gB_{ac^\p}(\x,\z)\gB_{bd^\p}(\x,\z)
           u(\x,\z) ,
\label{uabcd}
\end{equation}
and that (as $\x\rightarrow\Gamma$)
\begin{equation}
   v_{abc^\p d^\p}(\x,\z) = - \gB_a{}^{e^\p} \gB_b{}^{f^\p} R_{c^\p e^\p d^\p f^\p}(\z) +O(r).
 \label{vabcd}
\end{equation}
If $R_{cadb}u^c u^d \ne 0$ at the particle, then $\hB^\text{self}_{ab}$ is
non-differentiable there.

Similar to the electromagnetic field case, $\hB^\SSS_{ab}$ is an
inhomogeneous solution of \Deqn{EEqnHarm} satisfying the gauge
condition (\ref{deDonder}), if $T_{ab}$ is conserved.
 With the definition
\begin{equation}
  \hB^\R_{ab} \equiv \hB^\ret_{ab} - \hB^\SSS_{ab},
\end{equation}
$\hB^\R_{ab}$ is differentiable on $\Gamma$, and
\begin{equation}
  F^a = - \mu (g^{ab}+\dot z^a \dot z^b)\dot z^c \dot z^d
     (\nabla_{c} h^\R_{db} - \frac12 \nabla_b h^\R_{cd}) .
 \label{gravforce}
\end{equation}
Subject to this force, the particle moves along a worldline which is
actually a geodesic for a metric composed of the background geometry
complemented by $h^\R_{ab}$.
 While geodesic motion has been demonstrated in the past
\cite{Mino97,QuinnWald97}, only in our case is the reference metric,
$g_{ab}+h^\R_{ab}$, a vacuum solution of the Einstein equations through
$O(\mu)$.

\section{Discussion} For the clearest demonstration of the
impact of our analysis, we consider the free motion of a particle of small
mass $\mu$ in the purely gravitational case.
 With no a priori knowledge of the background geometry, an observer makes
local measurements of the metric within a neighborhood of the worldline of
the particle. That field has two distinct contributions.
 The first is the background metric combined with $h^\R_{ab}$ ---
this combination appears as an ``external'', homogeneous field; no
local measurement distinguishes $h^\R_{ab}$ from the background.
 The second comes from $h^\SSS_{ab}$; for free motion the observer knows
this contribution to be simply the $\mu/r$ part of the metric with its
tidal distortion from the ``external,'' homogeneous field\cite{Det01}.
 As a consequence of \Deqn{gravforce}, the local observer naturally sees
that the worldline of the particle is a geodesic in the combined,
``external'' homogeneous field which he measures. Making only local
measurements near the worldline, the observer sees no radiation, no
local source for the ``external'' field and also no effect which he
would be compelled to describe as radiation reaction.

 \acknowledgments{
We are grateful to Amos Ori and Robert Wald for comments on an
earlier version of this letter. This research has been supported in
part by NSF Grant No. PHY-9800977 (B.F.W.) and NASA Grant No.
NAGW-4864 (S.D.) with the University of Florida.}


\end{document}